\documentstyle[preprint,aps]{revtex}
\input epsf.tex
\def\DESepsf(#1 width #2){\epsfxsize=#2 \epsfbox{#1}}
\begin{document}

\preprint{\vbox{\hbox{OITS-620}\hbox{}}}
\draft
\title{Constraints from b$\rightarrow$ s $\gamma$ on gauge-mediated supersymmetry breaking models}
\author{\bf N. G. Deshpande, B. Dutta, and Sechul Oh }\address{ Institute of Theoretical Science,
University of Oregon, Eugene, OR 97403}
\date{November, 1996}
\maketitle
\begin{abstract}We consider the branching ratio of $b\rightarrow s\gamma$ in gauge mediated
supersymmetry breaking theories. Useful bounds on the parameter
space of these models are derived from the experimental bounds on
$b\rightarrow s\gamma$. Constraints on masses of NLSP are presented as a function of $\tan\beta$ and
$M/\Lambda$ for $\mu<0$ and $\mu>0$.
\end{abstract}

\pacs{PACS numbers:12.60.Jv,13.25.Hw, 12.10.Dm}

\newpage  There has been a tremendous recent interest in the  phenomenological implications of gauge
mediated supersymmetry breaking theories\cite{{DN},{pheno},{BBCT},{Martin},{Hotta}}. These
theories are characterized by gravitino  as the lightest supersymmetric particle(LSP) and have
signatures for supersymmetry which are distinct from the usual minimal supersymmetric standard
model(MSSM). The interest in these theories is especially
heightened because of the possible explanation of the peculiar
event seen at CDF with final state containing
$e^{+}e^{-}\gamma\gamma$\cite{SP} and missing transverse energy. A successful explanation of this
event requires the neutralino to be the next to lightest
supersymmetric particle (NLSP). Whether or not this explanation
withstands the test of time, it would seem important to examine in
detail the mass constraints on NLSP that ensue in these models
from the rare decay
$b\rightarrow s\gamma$ where SUSY contributions occur in one loop diagrams. Although some
preliminary work exists in special cases \cite{BBCT}, ours is the first comprehensive study in the
full allowed parameter space.

We shall analyze $b\rightarrow s\gamma$ in the gauge mediated SUSY breaking model in the whole range
of combinations of $\Lambda$ and $M$ (where $\Lambda$ is the effective SUSY breaking scale and $M$
is the messenger scale), the number
$n$ of ($n_5+{\bar n_5}$) pairs, and a range of $\tan\beta$ starting from small values to large
values. We take 
$\Lambda
\sim O(100 TeV)$ since soft SUSY breaking scalar masses are then of the order of the weak scale. The
parameter is
$M\ge\Lambda$. There could be a large hierarchy $M>>\Lambda$\cite{Hotta}, however the upper bound of
the gravitino mass
$\sim 1\, KeV$ restricts the $M/\Lambda <10^4 $\cite{PP}. The number of
 pairs $n$ is  restricted to 4 by the perturbative unification of the gauge coupling. The 10
representation can be included by noting that one
$n_{10}+{\bar n_{10}}$ pair corresponds to
$n$=3.  

The induced gaugino and scalar masses at the scale
M are \cite{comment}:
\begin{equation}
\tilde M_i(M) = n\,g\left({\Lambda\over M}\right)\, {\alpha_i(M)\over4\pi}\,\Lambda.
\label{bc1}
\end{equation}
\begin{equation}
\tilde m^2(M) = 2 \,(n)\, f\left({\Lambda\over M}\right)\,
\sum_{i=1}^3\, k_i \, C_i\,
\biggl({\alpha_i(M)\over4\pi}\biggr)^2\,
\Lambda^2
\label{bc2}
\end{equation}
where $k_i$ and  $C_i$ are 1,1,3/5 and 4/3,
3/4, and $Y^2$ for SU(3), SU(2), and U(1),
respectively.  The values of $C_i$ apply only to the fundamental representations of SU(3) and
SU(2) and are zero for the gauge singlets.  

We use the exact messenger-scale threshold functions \cite{Martin}

\begin{eqnarray} g(x)&=&{1+x\over x^2}\log(1+x)\;\;+\;\; (x\rightarrow-x)\ ,\\[2mm] f(x)&=&{1+x\over
x^2}\biggl[\log(1+x)-2{\rm Li}_2
\left({x\over1+x}\right) + {1\over2}{\rm Li}_2
\left({2x\over1+x}\right)\biggr]\;\; +\;\; (x\rightarrow-x)\ ,
\end{eqnarray}
rather than the limiting values, $f(0)=g(0)=1$ and $f(1)=0.7$ and $g(1)=1.4$.
  We shall require that electroweak symmetry be radiatively broken. We use
$\alpha_s=1.20$, ${\rm sin}^2\theta_w=0.2321$ and
$\alpha=1/127.9$ at the weak scale as the gauge coupling inputs. 
We first go up to the messenger scale $M$ with gauge and Yukawa
couplings, and fix the sparticle masses with the boundary
conditions (\ref{bc1}) and (\ref{bc2}). We next go down with the
6$\times$ 6 mass matrices for the squarks and sleptons to find the
sparticle spectrum for large as well as small
$\tan\beta$. We use the RGE's given in the reference \cite{VB}.  Also, we do not choose a particular
model for $\mu$. Note that the soft Higgs mass parameters
$m_{H_{1}}^2$ and
$m_{H_{2}}^2$ from eq.~(\ref{bc2}), along with the ratio of the vacuum expectation values
$\tan\beta(\equiv v_1/v_2)$ uniquely specifies
$\left|\mu\right|$, where $\mu$ is the coefficient of the bilinear Higgs term in the
superpotential.We will use two extreme values of
$\tan\beta$ equal to 3 and 42 (for $n=1$) for illustration. It is interesting that in the case of
$\tan\beta=3$ neutralino is the NLSP(next to lightest SUSY particle) for all values of 
$M>\Lambda$; however for
$\tan\beta=42$, either the stau or the neutralino is the NLSP depending on whether $\mu$ and
$M/\Lambda$ are small or large.  For
$n=2$, however stau can be NLSP even for the small $\tan\beta$ for the lower ratios of
$M/\Lambda$. When stau is the NLSP, these models can not explain ``the" event in Tevatron. However
the scenarios can have other interesting collider signatures both in Tevatron as well as LEP. 
 Calculation of $b\rightarrow s\gamma$ amplitude involves the coefficients of short distance
photonic and gluonic operators $c_7(M_w)$ and $c_8(M_w)$. Effects of QCD corrections to two loops is
then carried out. For the Standard Model these calculations are given in Ref.\cite{TI}. Recent
attempts to make partial corrections in the  next to leading order \cite{GH} do not indicate any
large terms. Contributions from various supersymmetric contributions are given in a generic form in
Ref.\cite{AM}. We use our calculated mass spectrum and the couplings to calculate
$b\rightarrow s\gamma$ rate. The results depend on $\Lambda$, $M$,
$\tan\beta$, $n$ and sign of $\mu$. We shall use more
physical variables $\tan\beta$,
$\mu$, sign of $\mu$, $M/\Lambda$ and $n$. Our figures are for the minimal case of $n=1$, and we
shall remark on the situation for higher $n$. The total amplitude has contribution from the W-loop,
charged Higgs (H$^{\pm}$) loop, chargino ($\chi^{\pm}$) loop, neutralino ($\chi^{0}$) loop and the
gluino (${\tilde g}$) loop. We find that the neutralino and the gluino contributions to the
amplitude are less than
$1\%$ in the whole range of parameter space. The charged Higgs contribution adds constructively to
the W-loop contribution. The chargino contribution can occur with either sign, but is generally much
smaller than the Higgs contribution. An exception is when $\tan\beta $ is large and $\mu<0$ and
chargino interference opens up the allowed parameter space.

We consider branching ratio vs $\mu$ for either sign of $\mu$ in two different scenarios: (a)
$\tan\beta=42$ and $M=1.1\Lambda$ and
$M=10^{4}\Lambda$; (b) $\tan\beta=3$ and $M=1.1 \Lambda$ and
$M=10^{4}\Lambda$ for n=1. The two different values of $M$ form the boundaries of the envelope of
parameter space that would be traced by any relation between $\Lambda$ and $M$. We exclude the case
$M=\Lambda$ since it produces a massless scalar in the messenger sector. We have used the particle
data values for the CKM matrix elements and also have imposed the constraint
$\left|V_{ts}^{*}V_{tb}\right|^2/\left|V_{cb}\right|^2=0.95\pm 0.04$
\cite{BM}.

In Figure 1a  corresponding to scenario (a) we  display $b\rightarrow s\gamma$ branching ratio as a
function of $\left|\mu\right|$ for $\mu>0$. Solid lines represent
$M=1.1 \Lambda$ and the dashed lines
$M=10^4 \Lambda$. CLEO bound $1\times 10^{-4}<Br(b\rightarrow s
\gamma)<4.2\times 10^{-4}$ at 95$\%$ CL clearly rules out the smaller values of $\mu$. We find
$\mu>720$. When the branching ratio is
$4.2\times 10^{-4}$ the NLSP (stau in the case of $M=1.1 \Lambda$) mass is 182 GeV. Fig 1b displays
the scenario(a) for
$\mu<0$. The chargino destructive interference for larger values of
$M/\Lambda$ does not yield any useful constraint on the parameter
space. In this parameter space, lighter mass NLSP solutions
correspond to stau as NLSP. The extreme left ends of the curves
correspond to a bound on the lightest slepton mass
$\sim$ 65 GeV. 

In Fig 2a, 2b we consider two cases in scenario (b), for
positive and negative
$\mu$ respectively. The left ends of the curves correspond to the lowest chargino mass bound which
we have taken to be around 75 GeV. In these cases, chargino contribution is very small. The
variation with M is also small. We then find that the CLEO constraint leads to
$\mu>460$ GeV for positive $\mu$ and
$\left|\mu\right|>426$ GeV for negative $\mu$ when $M=1.1 \Lambda$.

Constraints on $\mu$ can be translated into bounds for the masses of supersymmetric particles. We
are interested in displaying these bounds for the masses of NLSP. In Fig 3a. we display the lower
bound on neutralino mass as a function of $\tan\beta$ for $\mu>0$ and $\mu <0$ for two limiting
values of $M$, i.e. $M=1.1\Lambda$ and $M=10^4\Lambda$. Again as
$\tan\beta$ becomes larger, chargino interference in $\mu<0$ case for
$M=10^4 \Lambda$ removes any useful bounds. Though the variation of the ratio of $M/\Lambda$ does
not produce much difference in the branching ratio  for the lower values of $\tan\beta$, but it
has bigger effects in the bounds for the NLSP. In Fig 3b we display the lower bounds on  NLSP
when
$\tan\beta\ge 31$. Four cases considered are similar to Fig 3a. However, we do not have any
CLEO bound on the NLSP when $\mu<0$ and $M=10^4\Lambda$ throughout the range of $\tan\beta$
displayed. The solid curve and the dashed curve corresponds to the bounds on stau mass which
is the NLSP and the dot-dashed curve corresponds to the bounds on the neutralino mass, which
is the NLSP in this case. The bound on $\mu$, for the positive values of $\mu$ is a monotonic
function increasing from 460 GeV for
$\tan\beta=3$ to 720 GeV for $\tan\beta=42$ when
$M=1.1 \Lambda$. For $M=10^4\Lambda$, the bound increases from 480 GeV to 920 GeV in the same range
of $\tan\beta$. These large values of $\mu$ raise a problem of fine tuning.

For $n=2$, the constraint on the NLSP mass is higher, e.g. for
$\tan\beta=3$, lowest mass for the stau (since it is the NLSP) allowed by CLEO data would be 128 GeV
when $\mu<0$ and $M=1.1\Lambda$, however for
$M=10^{4}\Lambda$ the lowest mass allowed for the NLSP is 72 GeV and it is neutralino. Almost
the same thing happens for $\mu>0$, the lowest stau mass allowed is 136 GeV for
$M=1.1\Lambda$ and the lowest neutralino mass is 115 GeV for
$M=10^4\Lambda$. We also have the same conclusion for the $\tan\beta=42$, i.e the lowest mass
for the stau (since it is the NLSP) allowed by CLEO data would be  83 GeV when $\mu<0$ and
$M=1.1\Lambda$, however for
$M=10^{4}\Lambda$, there is no bound on the NLSP (stau), for
$\mu>0$, lowest stau mass(NLSP) allowed is  183 GeV for $M=1.1\Lambda$ and the lowest stau
mass (NLSP) is  201 GeV for
$M=10^4\Lambda$. We have assumed the value of 175 GeV for the top running mass. The results are
rather insensitive to this mass. A variation of 5 $\%$ in mass results in the change of branching
ratio of less than $1\%$.

In conclusion, we have used the CLEO bound on the branching ratio for
$b\rightarrow s\gamma$ to limit the parameter space of the gauge mediated supersymmetry breaking
models. We have found useful bounds on masses of NLSP. We also have found that for positive $\mu$,
irrespective of
$\tan\beta$, $\mu$ is restricted to large values. Since this raises the problem of fine tuning, our
analysis shows that gauge mediated model generally favors negative $\mu$ solutions. When
$\mu$ is negative the available parameter space increases with the ratio of
$M/\Lambda$. In the near future, with an improved bound on the branching ratio for
$b\rightarrow s\gamma$, it will be possible to put more severe constraints on the parameter
space.\\

This work was supported by Department of Energy Grant DE-FG03-96ER-40969.

\newpage
\begin{itemize}

\item[Fig. 1~:] {{Plots for $b\rightarrow s\gamma$ branching ratio as a
function of $\left|\mu\right|$ for $\tan\beta=42$.\\
Solid lines correspond to $M=1.1 \Lambda$.\\
dashed lines correspond to $M=10^4 \Lambda$.\\
a) for $\mu>0$\\ 
b) for $\mu<0$}\label{fig1}}. 
 
\item[Fig. 2~:] {{Plots for $b\rightarrow s\gamma$ branching ratio as a
function of $\left|\mu\right|$ for $\tan\beta=3$.\\
Solid lines correspond to $M=1.1 \Lambda$.\\
dashed lines correspond to $M=10^4 \Lambda$.\\
a) for $\mu>0$\\ 
b) for $\mu<0$}\label{fig2}}.  

\item[Fig. 3~:] {{3a) Bound on NLSP (neutralino) mass as a function of $\tan\beta$.\\ 
 Curves from the top (around $\tan\beta$=5) correspond to cases: (i) $\mu > 0$ and  
 $M=1.1 \Lambda$, (ii) $\mu < 0$ and $M=1.1 \Lambda$, (iii) $\mu > 0$ and 
 $M=10^4 \Lambda$, (iv) $\mu < 0$ and $M=10^4 \Lambda$.\\
 3b) Bound on NLSP (stau for the solid and dashed lines and neutralino for the 
 dot-dashed line) mass as a function of higher range of $\tan\beta$.\\ 
 Curves from the top (around $\tan\beta$=32) correspond to cases (i) $\mu > 0$ and  
 $M=1.1 \Lambda$, (ii) $\mu > 0$ and $M=10^4 \Lambda$, (iii) $\mu < 0$ and 
 $M=1.1 \Lambda$}\label{fig3}}.

\end{itemize}

\begin{figure}[htb]
\vspace{1 cm}

\centerline{ \DESepsf(tan42.epsf width 12 cm) }
\smallskip
\caption {}
\vspace{1 cm}

\centerline{ \DESepsf(tan3.epsf width 12 cm) }
\smallskip
\caption {}
\vspace{1 cm}

\centerline{ \DESepsf(NLSPmass.epsf width 12 cm) }
\smallskip
\caption {}
\vspace{3 cm}
\end{figure}

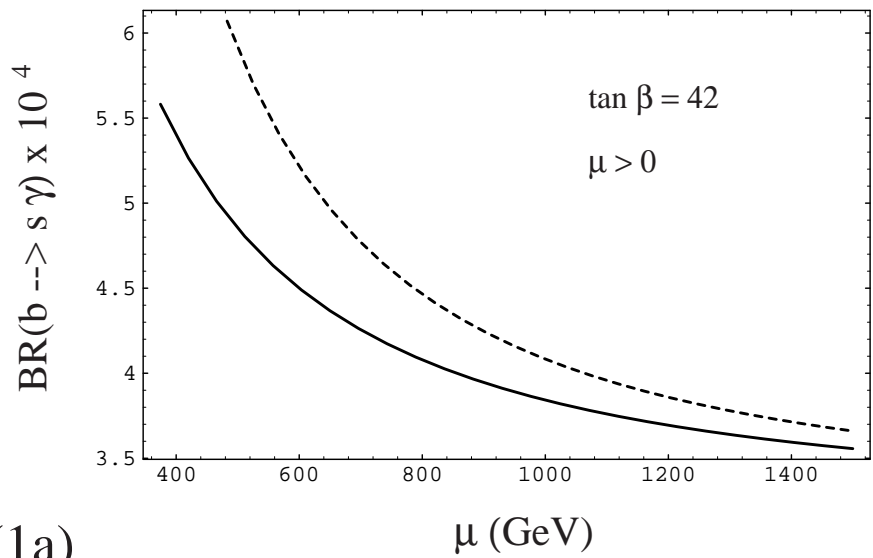
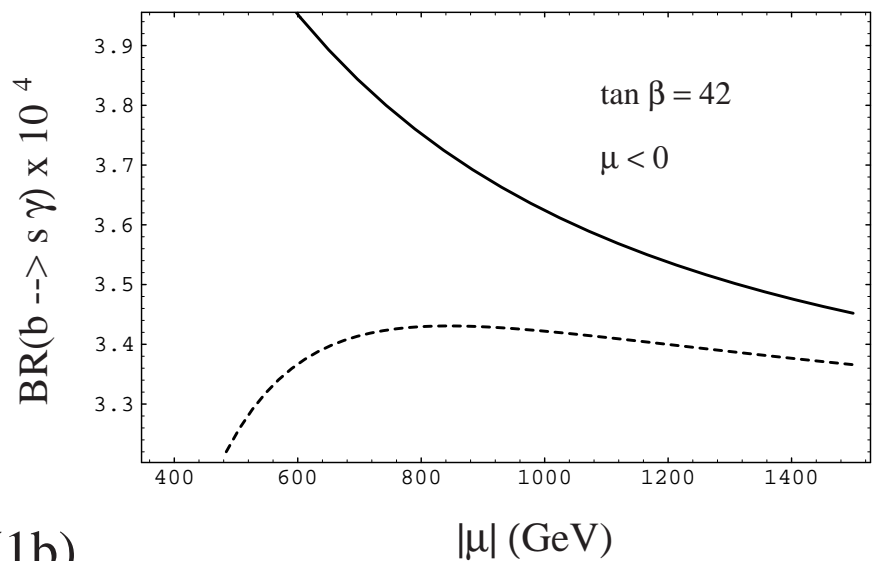
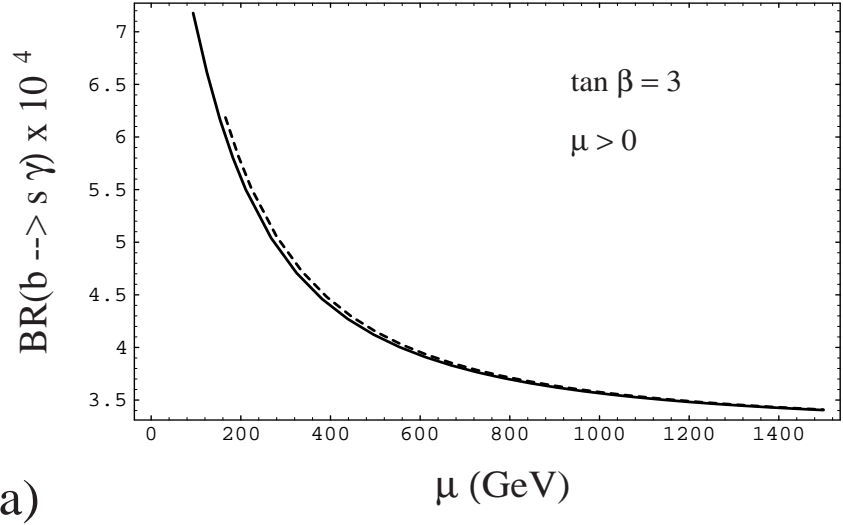
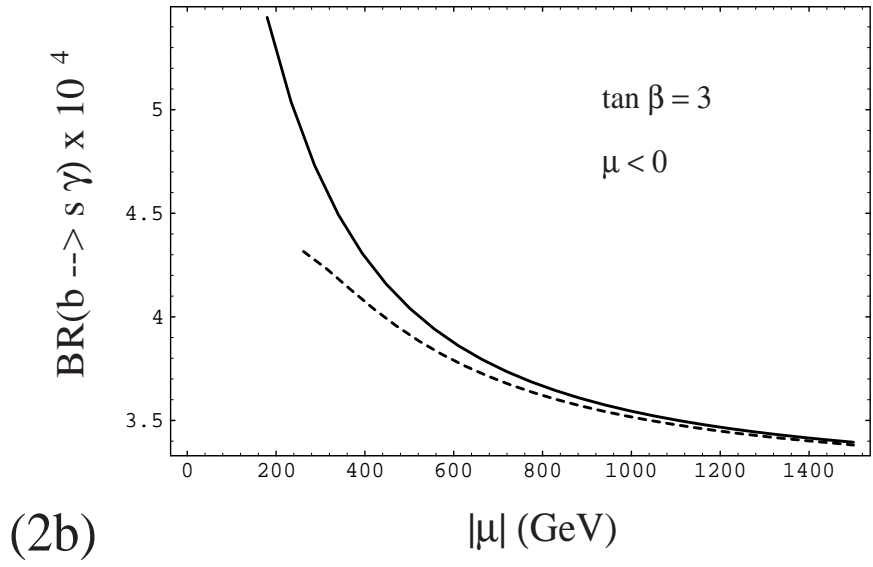
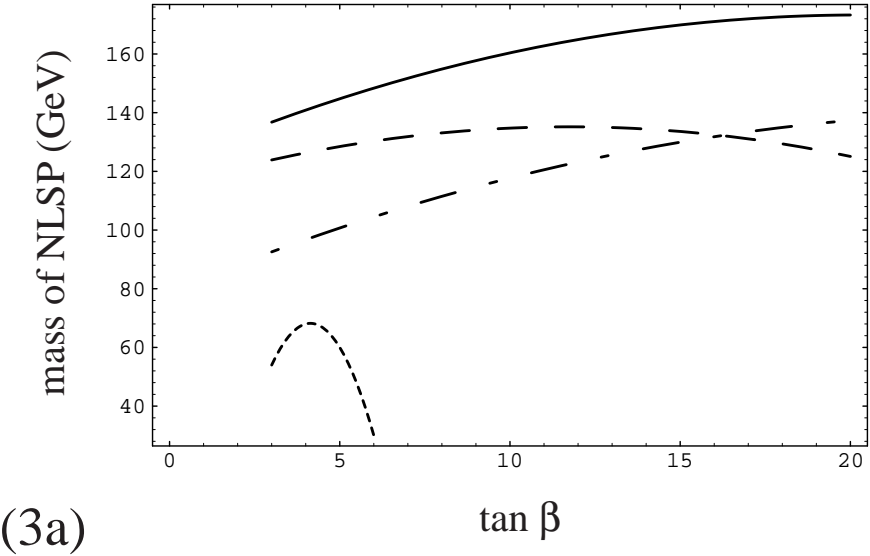
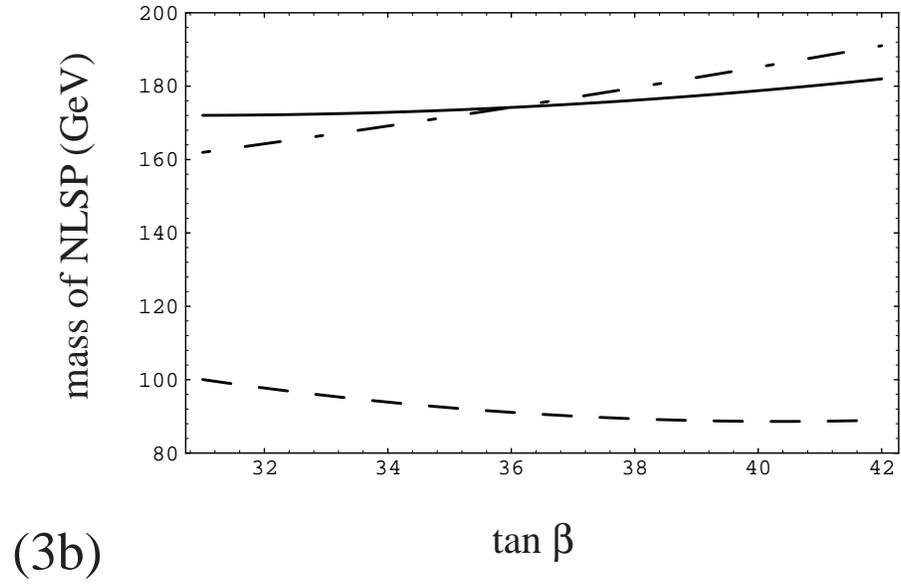
\end{document}